
-------------------------------------------------------------------------

\input phyzzx
%
%
%
%
\def\abstract{\vskip\frontpageskip
   \centerline{\bf {\titlestyle Abstract}} \vskip\headskip }
\def\ack{\par\penalty-100\medskip \spacecheck\sectionminspace
   \line{\hfil {\titlestyle {\bf Acknowledgements}} \hfil}
   \nobreak\vskip\headskip }
\def\refout{\par \penalty-400 \vskip\chapterskip
  \spacecheck\referenceminspace \immediate\closeout\referencewrite
  \referenceopenfalse
  \line{\hfil {\titlestyle {\bf References}} \hfil}\vskip\headskip
  \input \jobname.refs
  }
\def\figout{\par \penalty-400 \vskip\chapterskip
  \spacecheck\referenceminspace \immediate\closeout\figurewrite
  \figureopenfalse
  \line{\hfil {\titlestyle {\bf Figure Captions}} \hfil}
  \vskip\headskip
  \input \jobname.figs
  }
%
%
\def\NPrefs{\let\therefmark=\NPrefmark \let\therefitem=\NPrefitem}
\def\NPrefmark#1{[#1]}
\def\NPrefitem#1{\refitem{[#1]}}
\NPrefs
%
%
%
\unnumberedchapters
\Pubnum={TIT/HEP--173 \cr STUPP--91--120}
%
%
%
%
%
%
%
\REFS\brka{E. Br\'ezin and V.A. Kazakov,
           {\it Phys.\ Lett.\ }{\bf 236B} (1990) 144;
           M. Douglas and S. Shenker,
           {\it Nucl.\ Phys.\ }{\bf B335} (1990) 635;
           D.J. Gross and A.A. Migdal,
           {\it Phys.\ Rev.\ Lett.\ }{\bf 64} (1990) 127;
           {\it Nucl.\ Phys.\ }{\bf B340} (1990) 333.}
\REFSCON\gmil{D.J. Gross and N. Miljkovi\'c,
           {\it Phys.\ Lett.\ }{\bf 238B} (1990) 217;
           E. Br\'ezin V.A. Kazakov and A. Zamolodchikov,
           {\it Nucl.\ Phys.\ }{\bf B338} (1990) 673;
           P. Ginsparg and J. Zinn-Justin,
           {\it Phys.\ Lett.\ }{\bf 240B} (1990) 333;
           G. Parisi, {\it Phys.\ Lett.\ }{\bf 238B} (1990) 209, 213;
           J. Ambj\o rn, J. Jurkiewicz and A. Krzywicki,
           {\it Phys.\ Lett.\ }{\bf 243B} (1990) 373;
           D.J. Gross and I.R. Klebanov,
           {\it Nucl.\ Phys.\ }{\bf B344} (1990) 475;
           {\it Nucl.\ Phys.\ }{\bf B354} (1991) 459.}
\REFSCON\dika{J. Distler and H. Kawai,
           {\it Nucl.\ Phys.\ }{\bf B321} (1989) 509;
           J. Distler, Z. Hlousek and H. Kawai,
           {\it Int.\ J. of Mod.\ Phys.\ }{\bf A5} (1990) 391; 1093;
           F. David, {\it Mod.\ Phys.\ Lett.\ }{\bf A3} (1989) 1651.}
\REFSCON\seirev{N. Seiberg,
           {\it Prog.\ Theor.\ Phys.\ Suppl.\ }{\bf 102} (1990) 319.}
\REFSCON\poltalk{J. Polchinski,
           in: Strings '90, eds. R. Arnowitt et al,
           (World Scientific, Singapore, 1991) p.\ 62;
           {\it Nucl.\ Phys.\ }{\bf B357} (1991) 241.}
\REFSCON\kitarev{Y. Kitazawa,
           Harvard preprint HUTP--91/A034 (1991).}
\REFSCON\gouli{M. Goulian and M. Li,
           {\it Phys.\ Rev.\ Lett.\ }{\bf 66} (1991) 2051;
           A. Gupta, S. Trivedi and M. Wise,
           {\it Nucl. Phys.\ }{\bf B340} (1990) 475.}
\REFSCON\dfku{P. Di Francesco and D. Kutasov,
          {\it Phys.\ Lett.\ }{\bf 261B} (1991) 385.}
\REFSCON\kita{Y. Kitazawa,
           Harvard preprint HUTP--91/A013 (1991),
           {\it Phys.\ Lett.\ }in press.}
\REFSCON\sataco{N. Sakai and Y. Tanii,
            Tokyo Inst.\ of Tech.\ and Saitama preprint
            TIT/HEP--168, STUPP--91--116,
            {\it Prog.\ Theor.\ Phys.\ }in press.}
\REFSCON\dotsenko{V.S. Dotsenko,
            Paris preprint PAR--LPTHE 91--18 (1991).}
\REFSCON\taya{Y. Tanii and S. Yamaguchi,
            {\it Mod.\ Phys.\ Lett.\ }{\bf A6} (1991) 2271.}
\REFSCON\kostov{I. Kostov,
           {\it Phys.\ Lett.\ }{\bf 215B} (1988) 499; D.V. Boulatov,
           {\it Phys.\ Lett.\ }{\bf 237B} (1990) 202; S. Ben-Menahem,
           SLAC preprint SLAC--Pub--5262 (1990).}
\REFSCON\grklne{D.J. Gross, I.R. Klebanov and M.J. Newman,
           {\it Nucl.\ Phys.\ }{\bf B350} (1991) 621;
           D.J. Gross and I.R. Klebanov,
           {\it Nucl.\ Phys.\ }{\bf B352} (1991) 671;
           {\it Nucl.\ Phys.\ }{\bf B359} (1991) 3;
           G. Moore, Yale and Rutgers preprint YCTP--P8--91,
           RU--91--12 (1991);
           A.M. Sengupta and S.R. Wadia,
           {\it Int.\ J. of Mod.\ Phys.\ }{\bf A6} (1991) 1961;
           G. Mandal, A.M. Sengupta and S.R. Wadia,
           Princeton preprint IASSNS--HEP/91/8 (1991);
           J. Polchinski, Texas preprint UTTG--06--91 (1991).}
\REFSCON\dagr{U.H. Danielsson and D.J. Gross,
           Princeton preprint PUPT--1258 (1991).}
\REFSCON\polch{J. Polchinski,
           {\it Nucl. Phys.\ }{\bf B324} (1989) 123;
           {\it Nucl.\ Phys.\ }{\bf B346} (1990) 253;
           S.R Das, S. Naik and S.R. Wadia,
           {\it Mod.\ Phys.\ Lett.\ }{\bf A4} (1989) 1033;
           S.R. Das and A. Jevicki,
           {\it Mod.\ Phys.\ Lett.\ }{\bf A5} (1990) 1639.}
\REFSCON\sata{N. Sakai and Y. Tanii,
           {\it Int.\ J. of Mod.\ Phys.\ }{\bf A6} (1991) 2743;
           I.M. Lichtzier and S.D. Odintsov,
           {\it Mod.\ Phys.\ Lett.\ }{\bf A6} (1991) 1953.}
\REFSCON\berkl{M. Bershadsky and I.R. Klebanov,
           {\it Phys.\ Rev.\ Lett.\ }{\bf 65} (1990) 3088;
           {\it Nucl.\ Phys.\ }{\bf B360} (1991) 559.}
\REFSCON\polyakov{A.M. Polyakov,
           {\it Mod.\ Phys.\ Lett.\ }{\bf A6} (1991) 635.}
\REFSCON\dejero{K. Demeterfi, A. Jevicki and J.P. Rodrigues,
           Brown preprints Brown-HET--795 and 803, (1991).}
\refsend
\titlepage
\title{\bf Factorization and Topological States in $c \!\! = \!\! 1$
Matter Coupled to 2-D Gravity}
\author{Norisuke Sakai}
\address{Department of Physics, Tokyo Institute of Technology \break
         Oh-okayama, Meguro, Tokyo 152, Japan}
\andauthor{Yoshiaki Tanii}
\address{Physics Department, Saitama University \break
         Urawa, Saitama 338, Japan}
\abstract
Factorization of the $N$-point amplitudes in two-dimensional $c=1$
quantum gravity is understood in terms of short-distance
singularities arising from the operator product expansion of
vertex operators after the Liouville zero mode integration.
Apart from the tachyon states, there are
infinitely many topological states contributing
to the intermediate states.
\endpage
%
%
%
%
%
%
%
It is very important to understand the non-perturbative results of
matrix models \NPrefmark{\brka,\gmil} from the viewpoint of
the usual continuum approach of the two-dimensional quantum gravity,
i.e.\ the Liouville theory \NPrefmark{\dika -\kitarev}.
Recently correlation functions on the sphere topology have
been computed in the Liouville theory \NPrefmark{\gouli -\taya}.
These results are consistent
with those of matrix models \NPrefmark{\kostov -\dagr}.
So far only conformal field theories with central charge
$c \le 1$ have been successfully coupled to quantum gravity.
\par
The  $c = 1$ case is the richest and the most interesting.
It has been observed that this theory can be regarded effectively
as a critical string theory in two
dimensions, since the Liouville field zero
mode provides an additional ``time-like'' dimension besides the
obvious single spatial dimension given by the zero mode of the $c=1$
matter \NPrefmark{\polch}.
Since there are no transverse directions, the continuous (field)
degrees of freedom are exhausted by the tachyon field.
In fact, the partition function for the torus topology
was computed in the Liouville theory, and was found to give precisely
the same partition function as the tachyon field alone
\NPrefmark{\sata,\berkl}.
However, there are evidences for the existence of other
discrete degrees of freedom in the $c=1$ quantum gravity.
Firstly, the correlation functions obtained in the matrix model
exhibit a characteristic singularity structure \NPrefmark{\grklne}.
In the continuum approach of the Liouville theory, Polyakov has
observed that special states with discrete momenta and
``energies'' can produce such poles, and has called these operators
co-dimension two operators \NPrefmark{\polyakov}.
More recently, the two-loop partition function has been computed in a
matrix model and evidence has been noted for the occurrence of these
topological sates  \NPrefmark{\dejero}.
It is clearly of vital importance to pin down the role played by
these topological states as much as possible.
In the critical string theory, the particle content of the theory and
unitarity has been most clearly revealed through the factorization
analysis of scattering amplitudes.
On the other hand, the factorization and unitarity of the Liouville
theory has not yet been well understood.
\par
The purpose of this paper is to understand the factorization of
$c=1$ quantum gravity in terms of the short-distance singularities
arising from the operator product expansion (OPE) of vertex
operators.
Since we are interested in the short distance singularities, we consider
correlation functions on a sphere topology only.
We find that the singularities of the amplitudes can be
understood as short-distance singularities of two vertex operators
and that infinitely many discrete states contribute
to the intermediate states of the factorized amplitudes,
apart from the tachyon states.
These are the co-dimension two operators of Polyakov
\NPrefmark{\polyakov} and presumably are
topological in origin.
We have also explicitly constructed some of the topological states.
\par
%
%
%
%
%
Let us consider the $c=1$ conformal matter realized by a single
bosonic field (string variable) $X$ coupled to the two-dimensional
quantum gravity. After fixing the conformal gauge
$g_{\alpha \beta} = {\rm e}^{\alpha \phi} \hat g_{\alpha \beta}$
using the Liouville field $\phi$, the $c=1$ quantum gravity can be
described by the following action on a sphere
 \NPrefmark{\dika -\kitarev}
$$
S =
{1 \over 8\pi} \int d^2 z \sqrt{\hat g} \left(
\hat g^{\alpha \beta} \partial_\alpha X \partial_\beta X +
\hat g^{\alpha \beta} \partial_\alpha \phi \partial_\beta \phi
- Q \hat R \phi + 8 \mu \, {\rm e}^{\alpha \phi} \right),
\eqn\action
$$
where the parameters are given by
$Q = 2 \sqrt{2}, \, \alpha= - \sqrt{2}$ .
Since the correct cosmological term operator \NPrefmark{\seirev}
in the $c=1$ case should be $\phi \, {\rm e}^{\alpha\phi}$ rather than
${\rm e}^{\alpha\phi}$, the renormalized (correct) cosmological
constant $\mu_r$ is given by the following procedure:
one should replace $\mu$ and
$\alpha$ by $\mu_r/(2 \epsilon)$ and $(1-\epsilon)\alpha$ and take
the $\epsilon \rightarrow 0$ limit \NPrefmark{\dfku,\sataco}.
We have also set the ``Regge slope parameter'' $\alpha'=2$.
The gravitationally dressed tachyon vertex operator with momentum
$p$ has conformal weight $(1, 1)$:
$$
O_p = \int d^2z \sqrt{\hat g} \; {\rm e}^{i p X} \,
      {\rm e}^{\beta(p)\phi} , \qquad
\beta(p) = - \sqrt{2} + \; |p|.
\eqn\vertop
$$
We have chosen the plus sign in front of $|p|$, following the argument
of refs.\ \NPrefmark{\seirev,\poltalk}.
We see that the Liouville zero mode can be regarded as
an ``imaginary time'' and the exponent $\beta(p)$ as ``energy''.
\par
The $N$-point correlation function of the vertex operators
\vertop\ is given by a path integral
$$
\eqalign{
\VEV{O_{p_1} \cdots O_{p_N}}
= & \int {{\cal D} X {\cal D} \phi  \over V_{SL(2, {\bf C})}} \;\,
    O_{p_1} \cdots O_{p_N} \; {\rm e}^{- S} \cr
= & \, { \Gamma(-s) \over -\alpha}
    \int \prod_{i=1}^N \left[ \, d^2 z_i \, \sqrt{\hat g} \, \right]
    {1 \over V_{SL(2, {\bf C})}}
    \VEV{ \prod_{j=1}^N {\rm e}^{ip_j X(z_j)}}_X \cr
  & \times \VEV{  \left( {\mu \over \pi} \int d^2 w \,
    \sqrt{\hat g} \, {\rm e}^{\alpha \tilde \phi(w)} \right)^s
    \prod_{j=1}^N {\rm e}^{\beta_j \tilde \phi(z_j)}}_{\tilde\phi},
}\eqn\nptint
$$
where $V_{SL(2,\bf C)}$
is the volume of the $SL(2,{\bf C})$ group and
powers of the string coupling constant
$g_{\rm st}^{-2}$ for the sphere topology are omitted.
The expectation value with $\tilde\phi$ denotes the path integral
over the non-zero mode $\tilde\phi$ of the Liouville
field $\phi=\phi_0+\tilde \phi$, after the zero mode $\phi_0$
integration.
The defining formula for $s$ can be regarded as
``energy-momentum conservation''
$$
\sum_{j=1}^N {\bf p}_j + s {\bf q} + {\bf Q} =0,
\eqn\emocon
$$
where ${\bf p}_j = (p_j, - i \beta_j)$,
${\bf q} = (0,-i\alpha)$ and ${\bf Q}=(0,-iQ)$
are two-momenta for tachyons, ``cosmological terms'', and the source.
For a non-negative integer $s$, we can evaluate the non-zero mode
$\tilde \phi$ integral
by regarding the amplitude as a scattering amplitude
of $N$-tachyons and $s$ ``cosmological terms''.
After fixing the $SL(2, {\bf C})$ gauge invariance
($z_1=0, z_2=1, z_3=\infty$), an integral representation for
the $N$-tachyon amplitude is given by
$$
\VEV{ \prod_{j=1}^N O_{p_j} }
= \; 2\pi \delta \left( \sum_{j=1}^N p_j \right)
    {1 \over -\alpha} \Gamma(-s)  \tilde A (p_1, \cdots, p_N),
\eqn\tildea
$$
$$
\eqalign{
  \tilde A
= &  \left({\mu \over \pi}\right)^s
   \int \prod_{i=4}^N d^2 z_i \prod_{j=1}^s d^2 w_j
   \prod_{i=4}^N \left( |z_i|^{ 2 {\bf p}_1 \cdot {\bf p}_i}
    \, |1-z_i|^{2 {\bf p}_2 \cdot {\bf p}_i} \right)
   \prod_{4 \le i < j \le N}
    |z_i-z_j|^{ 2 {\bf p}_i \cdot {\bf p}_j} \cr
 &  \times
    \prod_{i=4}^N \prod_{j=1}^s
    |z_i-w_j|^{2 {\bf p}_i \cdot {\bf q}}
    \prod_{j=1}^s \left( |w_j|^{2 {\bf p}_1 \cdot {\bf q}}
    \, |1-w_j|^{2 {\bf p}_2 \cdot {\bf q}} \right)
    \prod_{1 \le j < k \le s} |w_j-w_k|^{2 {\bf q} \cdot {\bf q}}.
}\eqn\nptzint
$$
\par
In spite of the non-analytic relation \vertop\ between
energy $\beta$ and momentum $p$, we need to continue analytically
the formula into general complex values of momenta
in order to explore the singularity structure.
Hence we will define the tachyon to have positive (negative)
chirality if $(\beta+\sqrt2)/p =1(-1)$ irrespective of the actual
values of momentum \NPrefmark{\polyakov}.
It seems to us that the operators with $\beta<-\sqrt2$
in eq.\ \nptint\ are free from the trouble
noted in  \NPrefmark{\seirev,\poltalk} since $\phi_0$ has already
been integrated out.
The physical values of momenta are reached by analytic continuation
in $s$, since $s$ is related to other momenta through
energy-momentum conservation \emocon.
For generic physical values of momenta, one finds a finite result
for the $N$-tachyon amplitudes.
However, the result is different in different chirality
configurations, since the amplitude is non-analytic in momenta.
\par
If $p_1$ has negative chirality and the rest $p_2,\cdots,p_N$
positive chirality,
the amplitude is given by \NPrefmark{\gouli -\dotsenko}
$$
\tilde A (p_1, \cdots, p_N)
=  {\pi^{N-3} [\mu\Delta (-\rho)]^s \over \Gamma(N+s-2)}
   \prod_{j=2}^N \Delta(1 - \sqrt{2} p_j),
\eqn\nptfin
$$
where  $\Delta (x)=\Gamma(x)/ \Gamma(1-x)$.
The regularization parameter $\rho$ is given by $\rho = - \alpha^2 /2$
and is eventually set equal to $-1$ after the analytic continuation
(in the central charge $c$). We should replace the combination
$\mu\Delta(-\rho)$ by the renormalized cosmological constant $\mu_r$,
since the correct cosmological term is $\phi \, {\rm e}^{\alpha\phi}$.
The amplitude exhibits singularities at
$p_j=(n+1)/\sqrt2$, $n=0, 1, 2, \cdots$, but has no
singularities in other combinations of momenta contrary to the dual
amplitudes in the critical string theory.
These poles for $n = 1, 2, \cdots$ will be shown to correspond to
topological states as argued by several people
\NPrefmark{\grklne,\polyakov}.
\par
The amplitudes with one tachyon of positive chirality and the rest
negative are given by changing the sign of $p_j$.
On the other hand, if each chirality has two or more tachyons,
$\tilde A$ is finite for generic momenta
but has the factor $1/\Gamma(-s)$.
Hence $\tilde A$ vanishes for more than two tachyons in each chirality,
when we consider non-negative integer $s$ in the following.
This property has been explicitly demonstrated for the four- and
five-tachyon amplitudes
in the Liouville theory \NPrefmark{\dfku,\polyakov},
and has been argued to be a general
property using the matrix model \NPrefmark{\grklne}.
Therefore we take it for granted that the tachyon scattering amplitudes
$\tilde A$ vanish for non-negative integer $s$, unless
there is only one tachyon in either one of the chiralities.
Let us note that our assertion is consistent with the argument for
vanishing S-matrix in ref.\ \NPrefmark{\grklne}:
they absorbed the $\Delta(1 \pm \sqrt{2} \, p)$ factor in
the amplitude to a renormalization factor of vertex operators,
which becomes infinite if there is only one tachyon in either one
of the chiralities  \NPrefmark{\dfku -\sataco}.
Because of this infinite renormalization,
their renormalized amplitudes vanish
even if there is only one tachyon in either one of the chiralities.
\par
In order to understand the poles of the amplitudes in terms of
short-distance singularities in the OPE,
we shall consider the case of $s=$ non-negative integers
by choosing the momentum configuration appropriately.
These amplitudes at non-negative integer $s$ represent
so-called ``bulk'' or ``resonant''
interactions \NPrefmark{\grklne,\polyakov}.
Here we shall take the case of $s = 0$ for the $N$-tachyon
amplitude with only one negative chirality tachyon ($p_1$), and
examine the $s=$ positive integers case at the end.
\par
First we shall illustrate the origin of short-distance singularities
in the simplest context by expanding the integrand of the four tachyon
scattering amplitude with $s=0$
(we fix $z_2=0,z_3=1,z_4=\infty$ and call $z_1=z$)
$$
\eqalign{
\tilde A (p_1,\cdots,p_4)
& =  \int d^2 z \, |z|^{ 2 {\bf p}_1 \cdot {\bf p}_2}
     \, |1-z|^{ 2 {\bf p}_1 \cdot {\bf p}_3} \cr
& \approx \int_{|z| \le \epsilon}  d^2 z \, |z|^{-2\sqrt2 \, p_2}
          \left| \, \sum_{n=0}^{\infty}
          \biggl({\Gamma(1-\sqrt2 \, p_3) \over n! \,
          \Gamma(-\sqrt2 \, p_3-n+1)}\biggr) (-z)^n \, \right|^2 \cr
& \approx \sum_{n=0}^{\infty} {(-1)^n \over (n!)^2}
{\pi \over n+1-\sqrt2 \, p_2} \prod_{j=3}^4\Delta(1-\sqrt2 \, p_j).
}\eqn\fourptzint
$$
This shows that all the singularities in $p_2$ in the full amplitude
are correctly accounted for by these short-distance singularities
near $z_1\sim z_2$.
Furthermore we find that successive poles are due to
successive terms in the OPE
$$
:{\rm e}^{i{\bf p}_1 \cdot {\bf X}(z_1)}:
:{\rm e}^{i{\bf p}_2 \cdot {\bf X}(z_2)}: \,
\sim \sum_{n=0}^{\infty}\biggl({1 \over n!}\biggr)^2
|z_1-z_2|^{2{\bf p}_1 \cdot {\bf p}_2+2n} \,
:{\rm e}^{i{\bf p}_2 \cdot {\bf X}(z_2)}
\partial^n\bar\partial^n {\rm e}^{i{\bf p}_1 \cdot {\bf X}(z_2)}:,
\eqn\tacope
$$
with ${\bf X}=(X, \phi)$.
The pole at $p_2=1/\sqrt2$ ($n=0$) is due to the tachyon
intermediate state.
The higher level poles ($n\ge 1$) are due to the topological states
which we discuss shortly.
\par
We next examine short-distance singularities in amplitudes with
five or more tachyons.
By the same token, we consider the short-distance singularities due
to the OPE \tacope\ of two vertex operators $p_1$ and $p_2$ at
$z_1$ and $z_2$.
Because of the kinematical constraint, these singularities give poles
in $p_2$ at $(n+1)/\sqrt2, n=0,1,2,\cdots$.
The residues of these poles are given by kinds of dual amplitudes
with $N-2$ tachyons $p_3,\cdots,p_N$, and an intermediate particle
of two-momentum ${\bf p}={\bf p}_1+{\bf p}_2$ (Fig.\ 1).
Similarly to the four tachyon case, the $p_2=1/\sqrt2$ ($n=0$) pole
is due to the tachyon intermediate state with negative chirality.
In fact we find that the residue of the pole $p_2=1/\sqrt2$ is
precisely given by the $N-1$ tachyon amplitude with a single
(intermediate state) tachyon ${\bf p}$ having negative chirality
and the rest ${\bf p}_3,\cdots,{\bf p}_N$
having positive chirality
$$
 \tilde A(p_1, p_2, p_3, \cdots, p_N)
 \approx {\pi \over (N-3)(1-\sqrt2 p_2)}
 \tilde A(p, p_3, \cdots, p_N).
\eqn\tacpol
$$
This shows that the factorization is valid similarly to critical
string theory.
By symmetry, we can explain the lowest poles in each individual
momentum $p_j=1/\sqrt2$ as the tachyon intermediate state in the OPE
of $p_1$ and $p_j$.
\par
For higher level poles, we explicitly evaluate the residue of
the short-distance singularities
up to $p = 3 / \sqrt{2}$ and up to $N = 5$.
For instance, the five tachyon amplitude has short-distance
singularities at $p_2=2/\sqrt2$ and $3/\sqrt2$
$$
\tilde A (p_1, \cdots, p_5) \approx
\left[ - {\pi^2 \over 2(2 - \sqrt{2} \, p_2)}
+ {\pi^2 \over 8(3 - \sqrt{2} \, p_2)} \right]
\prod_{j=3}^5 \Delta(1-\sqrt2 \, p_j).
\eqn\singfive
$$
The residues of these poles in fact correctly reproduce the residues
of the poles in the full amplitude.
It is rather difficult to compute the short-distance singularities
explicitly to an arbitrary level except for the four-point amplitude
that we have already worked out in eq.\ \fourptzint.
Therefore we content ourselves with the computation of lower level
singularities in explicitly demonstrating that the singularities of
the amplitudes all come from the short-distance singularities
of $p_1$ and $p_j$.
\par
Since the short-distance singularities should come from terms in
the OPE, we next examine the operators
responsible for these singularities.
For higher level poles, it has been pointed out that there are only
null states at generic values of momenta
\NPrefmark{\polyakov,\grklne}.
However, there are exceptional values of momenta where the null states
degenerate and new primary states emerge as a result.
These new primary states are called co-dimension two
operators by Polyakov \NPrefmark{\polyakov}, and special states
or topological states by other people \NPrefmark{\grklne,\berkl}.
We can construct vertex operators for these topological states
in the following way.
The Virasoro generators $L_m$ for the $c=1$ quantum gravity are the
sum of the generators of the free scalar $X$ and those of the Liouville
field $\phi$.
The condition for the existence of the pole at level $n$ is given by
$$
{\bf p} \cdot ({\bf p} + {\bf Q})=2(1-n).
\eqn\polen
$$
We should note that there are two branches of the solution for the
condition \polen
$$
\beta=-\sqrt2\pm\sqrt{p^2+2n}.
\eqn\enen
$$
Although Seiberg has noted trouble with the lower sign due to the zero
mode $\phi_0$ \NPrefmark{\seirev,\poltalk}, we consider both cases
here, since we are considering
OPE of vertex operators consisting of non-zero mode $\tilde\phi$ only
($\phi_0$ integration gave the $s$ ``cosmological terms'').
We shall call the upper sign solution
S- (Seiberg) type and the lower A- (anti-Seiberg) type.
We should construct the field of conformal weight $(1, 1)$
by taking linear combinations of monomials of derivatives of ${\bf X}$
multiplied by ${\rm e}^{i{\bf p} \cdot {\bf X}}$.
For instance, at level $n=1$ we find only one field with weight
$(1, 1)$ at generic values of momentum, i.e.\ $p\not=0$
$$
V^{(1)}
= \, {\bf p} \cdot \partial {\bf X} \;
  {\bf p} \cdot \bar \partial {\bf X} \,
  {\rm e}^{i{\bf p} \cdot {\bf X}}
= - L_{-1} \bar L_{-1} {\rm e}^{i{\bf p} \cdot {\bf X}}.
\eqn\onenull
$$
The above state is clearly null.
\par
However, the situation changes at $p=0$.
For the S-type, the operator vanishes at $p=0$.
Therefore we can construct a new operator by a limit
$$
V_{(1,1)}= \lim_{p\rightarrow 0}{V^{(1)} \over p^2}
= \partial X \bar \partial X.
\eqn\gravopt
$$
We easily find that this field is primary and not null.
This kind of a peculiar operator exists only at a discrete momentum
and hence it is called co-dimension two.
This is precisely the ``graviton'', i.e.\  the first topological state
which gives rise to the pole at $p_2=2/\sqrt2$ in eq.\ \fourptzint.
As for the A-type at $p = 0$, we find that the $(1, 1)$ operator
condition
does not constrain the polarization tensor multiplying the operator
$\partial {\bf X} \, \bar\partial {\bf X} \, {\rm e}^{i{\bf p} \cdot
{\bf X}}$.
Hence we again obtain a new primary field
$$
V'_{(1,1)}= \partial X \bar \partial X{\rm e}^{-2\sqrt2 \phi}.
\eqn\agravopt
$$
At level two, we find two independent fields of weight one
for the holomorphic part.
The two fields for the holomorphic part are
$$
\eqalign{
V^{(2)} & = \left( L_{-2} + {3 \over 2} L_{-1}^2 \right)
            {\rm e}^{i {\bf p} \cdot {\bf X}}, \cr
V^{(3)} & = L_{-1} \left( { 1\over 4} i \left[
            ( 8 - {\bf p} \cdot {\bf Q} ) {\bf p} - 2 {\bf Q}
            \right] \cdot \partial {\bf X} \,
            {\rm e}^{i {\bf p} \cdot {\bf X}} \right).
}\eqn\twonull
$$
Both fields are null.
These two operators are linearly dependent at a special value
of the momentum and we obtain a topological state.
For instance, the (2,1) topological state of S-type is given by
$$
\eqalign{
V_{(2, 1)}
= & \lim_{p \rightarrow {1 \over \sqrt{2}}}
    {6 \sqrt{2} \over p - {1 \over \sqrt{2}}}
    ( V^{(2)} - V^{(3)} ) \cr
= & \, ( 13 \, \partial X \partial X - \partial \phi \partial \phi
    - 6 \, i \, \partial X \partial \phi - \sqrt{2} \, i \, \partial^2 X
    - \sqrt{2} \, \partial^2 \phi ) \,
    {\rm e}^{{1 \over \sqrt{2}} \, i \, (X - i \phi)}.
}\eqn\prmop
$$
We find exactly the same situation for the antiholomorphic part.
We can continue to explore $(1,1)$ operators at higher levels similarly.
We expect that these $(1,1)$ operators are null fields for generic
values of momenta, and that, at special values of momenta,
these null states are not linearly
independent, namely they degenerate.
Then we obtain a new primary state from a limit of an appropriate
linear combination of these null states.
We expect to have both S-type and A-type topological states.
\par
There are other procedures to obtain topological states.
These states were found to originate from the gravitational dressing
of the primary states in the $c=1$ conformal field theory which
create the null descendants at level $n$ \NPrefmark{\dagr, \berkl}.
The momentum $p$ of the initial primary state and the level $n$ are
specified by two positive integers $(r, t)$ and thus the energy
$\beta$ of the topological state is also given by
$$
p={r-t \over \sqrt2},\qquad
n=rt, \qquad
\beta={-2 \pm (r+t) \over \sqrt2}.
\eqn\rtstate
$$
The upper (lower) sign corresponds to the S-(A-)type solution.
We find that these operators with $(r,t)=(1,1)$ and $(2,1)$ differ from
our operators \gravopt\  and \prmop\  respectively, only by a certain
amount of null operators.
\par
The OPE suggests that there may be other short-distance singularities
in other combinations of momenta if one considers other combinations of
vertex operators approaching to the same point.
For instance, short-distance singularities corresponding to
$k$ vertex operators approaching each other, say $z_1, \cdots, z_k$,
should give poles in $p_2+\cdots+p_k$.
It is most convenient to fix reduced variables
$u_j = (z_j-z_2) / (z_1-z_2)$ $(j = 1, \cdots, k)$
to take the short-distance limit
$z_1 \rightarrow z_2$.
The amplitude exhibits short-distance singularities whose
residues are given by a product of two dual amplitudes (Fig.\ 2)
$$
\eqalign{
\tilde A(p_1, \cdots, p_N)
& \approx \, {1 \over V_{SL(2, {\bf C})}} \int_{|z_1-z_2|\le \epsilon}
  d^2 z_1 d^2 z_2
  \prod_{i=3}^k d^2 u_i \prod_{j=k+1}^N d^2 z_j \,
  |z_1 - z_2|^{{\bf p} \cdot {\bf p} + {\bf Q} \cdot {\bf p} - 4} \cr
 \times & \prod_{1 \le i < j \le k}
  |u_i - u_j|^{2 \, {\bf p}_i \cdot {\bf p}_j}
  \prod_{i=1}^k \prod_{j=k+1}^N \left|
  1 + {z_1 - z_2 \over z_2 - z_j} \, u_i
  \right|^{2 \, {\bf p}_i \cdot {\bf p}_j}  \cr
 \times & \prod_{i=k+1}^N |z_2 - z_i|^{2 \, {\bf p} \cdot {\bf p}_i}
  \prod_{k+1 \le i < j \le N}
  |z_i - z_j|^{2 \, {\bf p}_i \cdot {\bf p}_j}.
}\eqn\twodual
$$
The dual amplitude with the original variables
$z_i$ ($i = 2, k+1, \cdots, N$) has
$N-k$ positive chirality tachyons ${\bf p}_{k+1},\cdots,{\bf p}_N$
and the intermediate state particle ${\bf p}$ (right side blob in
Fig.\ 2), whereas
the dual amplitude with the reduced variables
$u_j$ ($j = 3, \cdots, k$) has the
intermediate state particle $-{\bf p}-{\bf Q}$ and $k$ tachyons
${\bf p}_1,\cdots,{\bf p}_k$ whose chiralities are
positive except ${\bf p}_1$  (left side blob in
Fig.\ 2).
If ${\bf p}$ is the intermediate state momentum flowing into the
right side blob, the corresponding momentum for the dual amplitude of
the left side blob can be regarded as $-{\bf p}-{\bf Q}$.
This implies that the chirality of the tachyon intermediate state is
the same for both dual amplitudes.
In the case of the intermediate topological state, the type (S or A)
of the intermediate state for one dual
amplitude turns out to be opposite to the other dual amplitude.
In the present case of pinching together the single negative chirality
tachyon with the positive chirality tachyons,
energy-momentum conservation \emocon\ dictates
that the intermediate state tachyon has negative chirality.
On the other hand,  the
tachyon amplitudes are non-vanishing only if a single tachyon has one
of the chiralities and the rest have opposite chirality.
Therefore the dual amplitude with the reduced variables vanishes
except when it is the three-point function ($k=2$).
This is precisely the case we have evaluated already in eq.\ \tacpol.
As for the intermediate topological states, kinematics
dictates that it is of S-type (A-type) for the dual amplitude with the
original variables (reduced variables).
The three-point function with the A-type topological state ($k=2$)
is nothing but the OPE coefficient \tacope\ which we have seen
non-vanishing.
Four- and more- point functions with the A-type topological state
$(k\ge 3)$ are more difficult to compute.
The topological state of level $n$ consists of a linear cmbination of
monomials of derivatives of $X$ multiplied by
a vertex operator ${\rm e}^{i{\bf p} \cdot {\bf X}}$.
Both the number of $\partial$ and the number of $\bar \partial$
should be $n$ for each monomial.
The two momentum ${\bf p}$ is given by
$((r-t)/\sqrt2, -i(-2-r-t)/\sqrt2)$ for the $(r,t)$ topological state
of type A.
If we do not specify the coefficients of the monomials, we obtain
an operator containing the $(r,t)$ topological state together with
certain amount of null states.
We have taken such an operator as a substitute for the $(r,t)$
topological state of A-type at the level $n=rt$, and have explicitly
evaluated the dual amplitude with the
topological state for the case of four-point
function.
We have found it to vanish.
This amplitude arises as the left side blob in Fig.\ 2 contributing
to the pole of level $n=rt$ in the case of $k=N-r-1=3$.
We conjecture in general that the A-type topological state
gives vanishing dual amplitude except for the three-point function
($k=2$).
This property is presumably related to the Seiberg's finding that only
the S-type is physical.
Only in the three-point dual amplitude ($k=2$),
we can simply regard the factor for the blob of particles pinched
together (left side blob in Fig.\ 2)
as the coefficient of the OPE rather than the dual
amplitude.
\par
Other possibilities are short-distance singularities from
the pinching of $k$ tachyons all with positive
chirality, say ${\bf p}_2, \cdots, {\bf p}_{k+1}$.
We again find that the chirality of the intermediate state tachyon is
negative.
Hence the dual amplitude with the original variables contains two
tachyons with negative chirality ${\bf p}_1,{\bf p}$ besides $N-k-1$
positive chirality tachyons ${\bf p}_{k+2},\cdots,{\bf p}_N$.
Hence the amplitude vanishes except
for the three-point function $(N=k+2)$.
The non-zero result of the three-point function gives rise to
poles in the momentum $p_N$ of
the single tachyon with the positive chirality.
We can regard these poles to be the same short-distance singularities
as obtained in $z_N \rightarrow z_1$.
Kinematics shows that the intermediate topological states
in this case is of A-type (S-type) for the dual amplitude with the
original variables (reduced variables).
According to our conjecture above, the dual amplitude with
the original variables again vanishes except
for the three-point function $(N=k+2)$.
The non-zero result can be interpreted in the same way as the tachyon
intermediate state.
\par
These observations explain why there are only sigularities in the
individual $p_j$, and none in any combinations of momenta, although the
factorization of the $N$-tachyon amplitudes is valid through the OPE as
we have seen.
\par
Let us finally discuss the case of $s=$  positive integer.
The amplitudes with $s=$ positive integer can be obtained
from the $s = 0$ case as follows: we consider the $N+s$
tachyon scattering amplitude and take a limit of vanishing
momenta for $s$ tachyons and multiply by $(\mu/\pi)^s$.
There is one subtlety: at the vanishing
momenta, the chirality is ill-defined  \NPrefmark{\dfku -\sataco}.
We define the vanishing momenta taking
limit from the positive chirality tachyon.
In the limit, we obtain an $s$-th power of a singular factor
$\mu \Delta(0)$, which should be replaced by the renormalized
cosmological constant $\mu_r$.
In this way we find that the short-distance singularities
of the amplitudes with non-vanishing $s$ can be obtained correctly
once the short-distance singularities in the $s = 0$ amplitude are
correctly obtained.
Using the previous argument, we find that the only non-vanishing
short-distance singularities are from the OPE
of two vertex operators for tachyons.
Short-distance singularities from one or more ``cosmological term
operators'' approaching the tachyon vertex operators give a
vanishing value for the residue.
\par
\vskip 2mm
\ack
One of the authors (NS) thanks Y. Kitazawa and D. Gross
for a discussion on the Liouville theory.
We would like to thank Patrick Crehan for a careful reading
of the manuscript.
This work is supported in part by Grant-in-Aid for Scientific Research
from the Ministry of Education, Science and Culture (No.01541237).
\vskip 2mm
\refout
\vskip 2mm
\FIG\a{The factorization of the $N$-tachyon amplitude by the OPE
       of the operators 1 and 2. The signs $+$ and $-$ denote
       the chirality of the tachyons.}
\FIG\b{The factorization of the $N$-tachyon amplitude by the OPE
       of the operators $1, \cdots, k$. The signs $+$ and $-$ denote
       the chirality of the tachyons.}
\figout
\end